\documentclass[aps,pra,reprint,twocolumn,superscriptaddress]{revtex4-1}
\usepackage[utf8]{inputenc}
\usepackage{graphicx}
\usepackage[T1]{fontenc}
\usepackage[english]{babel}
\usepackage[top=2cm, bottom=2cm, left=3cm, right=3cm]{geometry}
\usepackage{lmodern}
\usepackage{verbatim}
\usepackage{amsmath}
\usepackage{amssymb}
\usepackage{amsthm}
\usepackage{libertine}
\usepackage{upgreek}
\usepackage{array,multirow,makecell}
\usepackage{color}
\usepackage[all]{xy}
\usepackage{tikz}
\usepackage{stmaryrd}
\tikzset{math3d/.style={x= {(-0.7cm,-0.2cm)}, z={(0cm,1cm)},y={(1cm,0cm)}}}
\usetikzlibrary{arrows}
\usetikzlibrary{decorations.markings}
\tikzset{
  myarrow/.style={
    decoration={markings,mark=at position 1 with {\arrow[scale=2,>=stealth]{>}}},   postaction={decorate}}}
\usepackage[e]{esvect}

\setcellgapes{1pt}
\makegapedcells
\newcolumntype{R}[1]{>{\raggedleft\arraybackslash }b{#1}}
\newcolumntype{L}[1]{>{\raggedright\arraybackslash }b{#1}}
\newcolumntype{C}[1]{>{\centering\arraybackslash }b{#1}}

\newcommand{\iin}{\mathrm{in}}
\newcommand{\out}{\mathrm{out}}
\newcommand{\tz}{\tilde{z}}

\newcommand{\tpsi}{\tilde{\varPsi}}

\newcommand{\bz}{\mathbf{z}}
\newcommand{\bF}{\mathbf{F}}
\newcommand{\bE}{\mathbf{E}}
\newcommand{\bV}{\mathbf{V}}
\newcommand{\br}{\mathbf{r}}
\newcommand{\bk}{\mathbf{k}}
\newcommand{\bpsi}{\mathbf{\Psi}}

\newcommand{\cS}{\mathcal{S}}
\newcommand{\cT}{\mathcal{T}}

\newcommand{\tA}{\tilde{\mathcal{A}}}

\begin{document}

\title{Improved Effective Range Expansion for Casimir-Polder potential}

\author{P.-P. Cr\'epin} \email[]{pierre-philippe.crepin@lkb.upmc.fr}
\affiliation{Laboratoire Kastler Brossel (LKB), Sorbonne Universit\'e, CNRS, 
ENS-PSL Universit\'e, Coll\`ege de France, Campus Pierre et Marie Curie, 75252, Paris, France}
\author{R. Gu\'erout} \email[]{romain.guerout@lkb.upmc.fr}
\affiliation{Laboratoire Kastler Brossel (LKB), Sorbonne Universit\'e, CNRS, 
ENS-PSL Universit\'e, Coll\`ege de France, Campus Pierre et Marie Curie, 75252, Paris, France}
\author{S. Reynaud} \email[]{serge.reynaud@lkb.upmc.fr}
\affiliation{Laboratoire Kastler Brossel (LKB), Sorbonne Universit\'e, CNRS, 
ENS-PSL Universit\'e, Coll\`ege de France, Campus Pierre et Marie Curie, 75252, Paris, France}
\date{\today}

\begin{abstract}
We study the effective range expansion of scattering on a real Casimir-Polder potential. We use Liouville transformations which transform the potential landscape while preserving the reflection and transmission amplitudes. We decompose the scattering calculation in two more elementary problems, one for the homogeneous $1/z^4$ potential and the other one for the correction to this idealization. We use the symmetries of the transformed problem and the properties of the scattering matrices to derive an \textit{improved effective range expansion} leading to a more accurate expansion of scattering amplitudes at low energy. 
\end{abstract}

\maketitle

\section{Introduction}

The \textit{effective range theory} was developed long ago  \cite{Hamermesh1947,Blatt1949,Bethe1949,Barker1949,Teichmann1951,Brueckner1951,Salpeter1951}
for studying nucleon scattering with scattering amplitudes described at low energies by a small number of parameters, 
namely a scattering length and an effective range in the simplest cases. In the present paper, we focus our attention on the study of atom scattering on the Casimir-Polder potential above a surface 
\cite{Lennard-Jones1936III,Lennard-Jones1936IV,Casimir1946,Casimir1948}.
The Casimir-Polder potential varies rapidly in the vicinity of the surface, which leads to quantum reflection even though the potential well is attractive \cite{Yu1993,Berkhout1993,Carraro1998,Shimizu2001,Druzhinina2003,Pasquini2004,Oberst2005,Zhao2008}. 

Quantum reflection should play a key role in the GBAR experiment that will test the weak equivalence principle on antihydrogen atoms \cite{Debu2012,Indelicato2014,Perez2015,Mansoulie2019} as it prevents the detection of antihydrogen atoms through annihilation on the detector \cite{Voronin2005jpb,Voronin2011,Voronin2012pra}. It can also be turned into a tool for reaching better experimental accuracy \cite{Dufour2014shaper,Dufour2015ahep}.  In a new quantum measurement methods recently proposed to improve the accuracy of GBAR experiment \cite{Crepin2019}, a precise knowledge of the Casimir-Polder shifts of quantum gravitational states \cite{Nesvizhevsky2002,Nesvizhevsky2005} is required, and this requirement can be met by mastering the effective range expansion for scattering amplitudes at low energy \cite{Crepin2017}.

We explain in the next section the motivation for building an improved effective range expansion in the case of real Casimir-Polder potentials. We then propose a new derivation based on Liouville transformations of the Schr\"odinger equation which map the physical potential landscape into another potential more convenient to study quantum reflection. We use symmetries and composition properties of scattering amplitudes to derive an improved effective range expansion, leading to more accurate predictions at low energy and opening new perspectives for precise spectroscopic measurement of quantum gravitational states.

\section{Motivations for an improved effective range expansion}

An (anti)hydrogen atom of mass $m$  at height $z$ above a horizontal surface is submitted to the gravity potential and
the Casimir-Polder potential which dominates at the distances considered, of the order of a few micrometers. The Casimir-Polder potential $V(z)$ is attractive at all distances, with characteristic asymptotic power laws, the so called \emph{Van der Waals} limit near the surface for $z\to0$ and \emph{retarded} limit far from it $z\to+\infty$~:
\begin{equation}
\begin{split}
\label{powerlaws}
&V(z) \simeq  -C_3/z^3 \equiv V_3(z) \quad,\quad z\to0^+~,\\
& V(z) \simeq -C_4/z^4 \equiv V_4(z) \quad,\quad z\to+\infty~.
\end{split}
\end{equation}
The long-range length scale corresponding to this potential is $\ell_4=\sqrt{2mC_4}/\hbar$ while the short-range length scale is $\ell_3=2mC_3/\hbar^2$. Typical values for helium and silica surfaces are given in table \ref{lengths}.

The scattering amplitudes for an atom falling down onto the surface can be calculated by solving the one-dimensional stationary Schr\"odinger equation obeyed by the wave function \cite{Berry1972,Friedrich2004}~:
\begin{equation}
\begin{split}
\label{base}
\psi''(z)+F(z)\psi(z)=0 \\
F(z)\equiv\frac{2m(E-V(z))}{\hbar^2}.
\end{split}
\end{equation}
The equation \eqref{base} can be solved numerically by imposing absorbing boundary conditions $\psi(0)=0$ on the antihydrogen atom, annihilated at contact on the surface. The potential vanishes at large distances and the wave function is written ($E=\hbar k^2/(2m)$ is the energy and $k$ the asymptotic wavevector)~:
\begin{equation}
\psi(z) \underset{z\to+\infty}{\sim} e^{-ikz}+r(k)~e^{ikz} ~.
\label{asymptote}
\end{equation}
The reflection amplitude $r(k)$ which describes the quantum reflection of the atom on the Casimir-Polder potential has been computed for antihydrogen falling down on different surfaces \cite{Dufour2013qrefl,Dufour2013porous,CrepinEPL2017}. 
It goes to $-1$ at the limit of low energies, with an asymptotic approach to this limit described by a  \textit{scattering length}~:
\begin{equation}
\tA(k) = -i\frac{1+r(k)}{1-r(k)} \quad,\quad
\underset{k\to0}{\lim} \frac{\tA(k)}{k} =  -i\ell. 
\end{equation}

For the ideal homogeneous potential $V_4$, the length $\ell$ coincides with the length scale $\ell_4$ \cite{OMalley1961,Arnecke2006}, that is real. This is why we chose to call $\ell$ the scattering length while the traditional scattering length -- often denoted by $a\equiv -i\ell$ -- is purely imaginary in that case. 
 For a real Casimir-Polder potential in contrast, it is not the case though it might have been expected that the low-energy behavior of $r(k)$ is determined by the long-range part of the potential. As shown in table \ref{lengths}, the scattering length computed for real potentials with antihydrogen on helium or silica surfaces significantly differs from this expectation.

\begin{table}[ht]
\centering
\begin{tabular}{c|c|c|c|}
\cline{2-4}
 & $\ell_4$ & $\ell$ & $\ell_3$   \\
\hline
\multicolumn{1}{|c|}{He} &~ $75.51$~ & ~$44.78-34.90~i$~ & ~$16.54$ ~ \\
\hline
\multicolumn{1}{|c|}{SiO$_2$} & ~$194.7$ &~ $~272.7-77.04~i ~ $  & ~$321.3$~ \\
\hline
\end{tabular}
\caption{Typical length scales $\ell_4$, $\ell_3$ and scattering length $\ell$ calculated for the real Casimir-Polder potential between antihydrogen and helium or silica surfaces (atomic units).}
\label{lengths}
\end{table}

Not only the scattering length but also the Taylor expansion of $\mathcal{A}(k)$ at low energies is required to compute the precise positions of gravitational quantum states \cite{Crepin2017}. 
This effective range expansion is known  for the ideal $V_4$ potential \cite{OMalley1961,Arnecke2006}. The parameters of the expansion  can be modified when assuming that the difference between the real potential $V$ and the long-range limit $V_4$  is a short-range potential. However this assumption is not valid for the real Casimir-Polder potential, where  the difference $V-V_4$ is known to behave essentially as another long-range potential $V_3$.
 
In the following, we will introduce Liouville transformations of the Schr\"odinger equation which change the potential landscape while exactly preserving reflection amplitudes \cite{Dufour2015jpb,Dufour2015epl}. 
We will see that this allows one to bypass the mathematical intricacies in the modified effective range theory for Casimir-Polder potentials, and then to get an improved asymptotic expansion of reflection amplitudes at low energies.

\section{Liouville transformations}

We consider now Liouville transformations which transform the potential and remove its divergence in the vicinity of the surface \cite{Dufour2015jpb,Dufour2015epl}. 

We first introduce the WKB phase~:
\begin{equation}
\phi_{dB}(z)=\int_{z_\phi}^z \sqrt{F(z')}dz' ~,
\end{equation}
where $z_\phi$ is a reference point to be fixed by choosing a phase origin for the wave function $\phi$ at  $z \rightarrow +\infty$ where the WKB phase is linear~:
\begin{equation}
\lim\limits_{z \rightarrow +\infty} (\phi_{dB}(z)-kz)=\phi.
\end{equation}
We define the Liouville transformation as a related change of coordinate and wavefunction  scaling ~:
\begin{equation}
\begin{split}
\label{gauge}
\bz&=\frac{\phi_{dB}(z)}{\sqrt{k\ell_4}}\\
\bpsi(\bz)&=\sqrt{\bz'(z)}~\psi(z) ~.
\end{split}
\end{equation}
The phase $\phi$ now corresponds to a translation length $\bz_\phi$
 in the new coordinates to be adjusted by symmetry considerations 
discussed later on~:
\begin{equation}
\bz_\phi \equiv \int_{z_\phi}^{z_0}\sqrt{\frac{F(z')}{k\ell_4}}~dz' ~.
\end{equation}

We can now rewrite the Schr\"odinger equation in the transformed coordinates~:
\begin{equation}
\begin{split}
\label{schbold}
\bpsi''&(\bz)+\bF(\bz)\bpsi(\bz)=0 ~,\\
&\bF(\bz)=\bE-\bV(\bz)~,
\end{split}
\end{equation}
with $\bF(\bz)$ given by a Schwarzian derivative  $\{\bz,z\}$~:
\begin{equation}
\begin{split}
\bF(\bz)&=\frac{F(z)-\frac{1}{2}\{\bz,z\}}{\bz'(z)^2}\\
\{\bz,z\}&=\frac{\bz'''(z)}{\bz'(z)}-\frac{3}{2}\frac{\bz''(z)^2}{\bz'(z)^2}\equiv2F(z)Q(z)~.
\end{split}
\end{equation} 
A significant \textit{badlands function} $Q(z)$ indicates  the \textit{badlands} where the WKB approximation fails and quantum reflection occurs. The transformed energy, potential and wave-vector can also be written~:
\begin{equation}
\begin{split}
\bE&=k\ell_4 ~,\\
\bV(\bz)&=k\ell_4\,Q(z)~,\\
\label{k}
\bk&=\sqrt{k\ell_4}~.
\end{split}
\end{equation}

For the ideal homogeneous potential $V_4(z)$ the transformed $\bV_4(\bz)$ reaches a maximum at $\bz_M$ \cite{Dufour2015jpb}~:
\begin{align}
&\bz_M\equiv \bz(z=\zeta)=\bz_{*}+\bz_\phi~,\quad \zeta\equiv \sqrt{\frac{\ell_4}k} \\
&\bz_{*}\equiv2\left[ _2F_1\left(\tfrac{1}{2},-\tfrac{1}{4};\tfrac{3}{4};-1\right)-\frac{1}{\sqrt{2}} \right]
=\frac{1}{\sqrt{\pi}}\Gamma \left(\tfrac{3}{4}\right)^2~, \nonumber
\end{align}
where $_2F_1$ is the hypergeometric function that relates $\bz$ and $z$ for homogeneous potentials \cite{Dufour2015}. 
In order to make the potential $\bV_4$ an even function of $\bz$, we chose $\bz_M=0$, that is also $\bz_\phi=-\bz_{*}$. 

\begin{figure}[ht]
   \center
   \includegraphics[width=\linewidth]{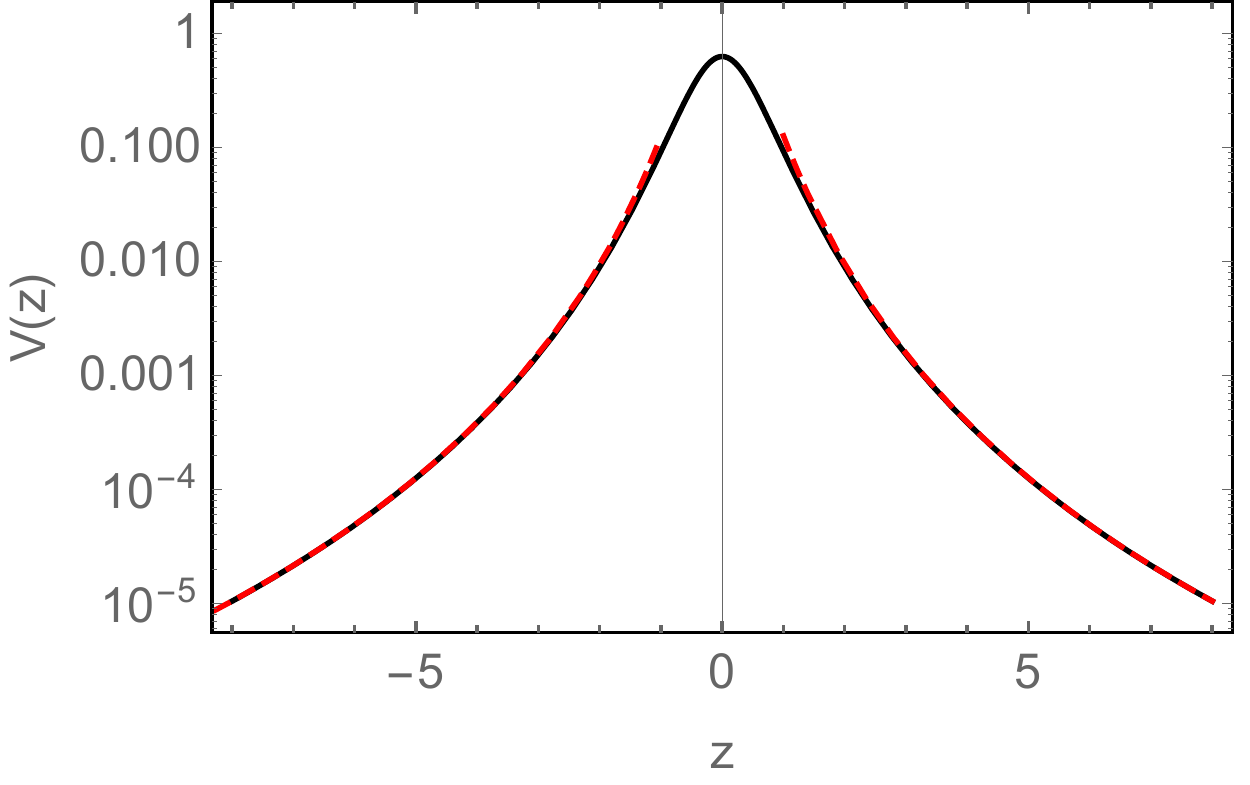}
   \caption{Black line represents the potential $\bV_4(\bz)$, an even function after Liouville transformation; the dashed red lines show its asymptotic behavior (colors on line).}
   \label{V4}
\end{figure}

The potential $\bV_4$ shows universal properties which do not depend on the amplitude $C_4$ of the $V_4$ potential after the change of coordinates is done as in  \eqref{gauge}. It can be rewritten as~:
\begin{equation}
\begin{split}
\bV_4(\bz)&=\frac{5}{8\cosh^3(2u)} \quad,\quad u\equiv \ln \frac{z}{\zeta} ~,\\
\bz&=\int_0^{u}\sqrt{2\cosh(2u')}\text{d}u' ~,
\label{z_u}
\end{split}
\end{equation}
 and its asymptotic behavior deduced~:
\begin{equation}
\bV_4(\bz) \underset{\bz\to \pm \infty}{\simeq} \frac{5}{\bz^6}
\quad,\quad \bz \underset{u\to \infty}{\simeq} e^{u}~.
\end{equation}
The new potential in Liouville coordinates is plotted as the black full line in figure \ref{V4}
with its  asymptotic behavior shown as dashed red lines.

\begin{figure}[ht]
   \centering
   \includegraphics[width=\linewidth]{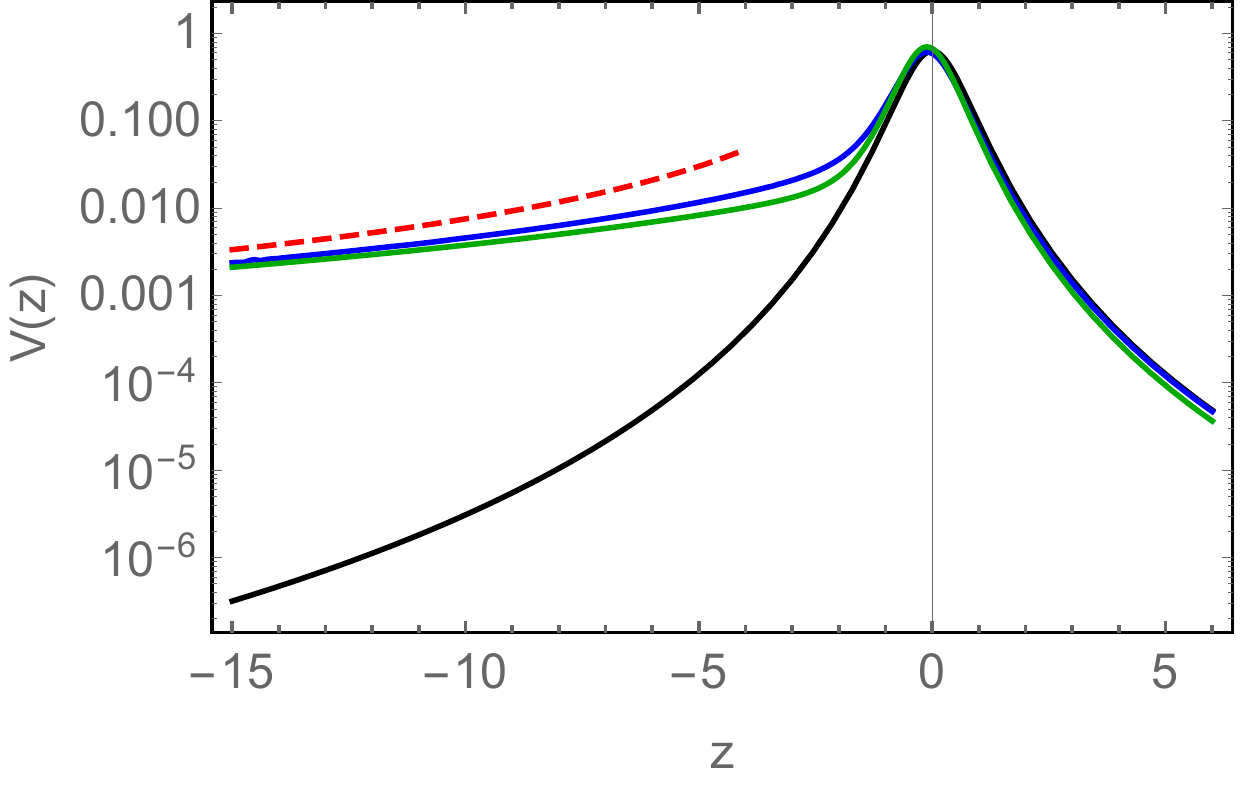}
   \caption{Casimir-Polder potentials plotted after Liouville transformation (blue line for He and green line for SiO$_2$), calculated at energy $E=1$~neV. The black line shows the $\bV_4$ potential for comparison while the red dashed line is the asymptotic behavior for the $\bV_3$ tail.}
   \label{V}
\end{figure}

The real Casimir-Polder potential $V$ departs from the ideal form $V_4$ at not too large distances, which breaks the symmetry and modifies the transformed $\bV$ after Liouville transformation, as can be seen in figure \ref{V}. The change
is particularly significant in the left part $\bz<0$ which corresponds to $z<\zeta$.

For $\bz \rightarrow -\infty$, $\bV$ behaves asymptotically, as $\bV_3$ and the expression of the latter potential is known (as for all homogeneous potentials $\bV_n$; see \cite{Dufour2015})~:
\begin{align}
&\bV_3(x)=3x\frac{1+16x^3}{16(1+x^3)^3} ~, \label{Bold} \\
&\bz=3x\left[_2F_1\left(\tfrac{1}{2},-\tfrac{1}{3};\tfrac{2}{3};-\tfrac{1}{x^3}\right)-\frac{2}{3}\sqrt{1+\frac{1}{x^3}}\right]~.
\nonumber
\end{align}
From \eqref{Bold}, we deduce the asymptotic behaviour of $\bV$ shown as the dashed red line in figure \ref{V}~:
\begin{align}
&\bV(\bz)\underset{\bz\rightarrow -\infty}{\simeq}\bV_3(\bz)\underset{\bz\rightarrow -\infty}{\simeq}\frac{3}{4\bz^2}~,
\\ &\bz \underset{x\rightarrow 0}{\simeq} -2/\sqrt{x}~. \nonumber
\end{align}

\section{Two-step scattering process}
After the Liouville transformation, the potential landscape is much smoother than the original one so that we can decompose the scattering process into two steps. The first step is the reflection on the universal long-range part $\bV_4$ felt by the atom when he falls down from large distances, while the second one is the reflection on the inner $\bV$ differing from $\bV_4$ when the atom has been transmitted trough the first barrier. 

The \textit{scattering matrix} $\cS$ connects the amplitudes of waves propagating \textit{out} and \textit{in}
and it has a general form for  one-channel scattering~:
\begin{equation}
\left(
\begin{matrix}
 a_{+}^\out\\
 a_{-}^\out
\end{matrix}
\right)
=\cS
\left(
\begin{matrix}
 a_{+}^\iin\\
 a_{-}^\iin
\end{matrix}
\right)
\quad,\quad
\cS=
\left(
\begin{matrix}
\overline{t}& r \\
\overline{r} & t
\end{matrix}
\right)~.
\end{equation}
It is also useful to define the \textit{transfer matrix} $\cT$ that relates left and right waves.
$\cS$ and $\cT$ matrices are related by an operation \cite{Genet2003}
defined for all matrices with a non-zero coefficient $m_{1,1}$~:
\begin{equation}
\begin{array}{lrcl}
  \Pi :
      &\left(
\begin{matrix}
m_{11}& m_{12} \\
m_{21} & m_{22}
\end{matrix}
\right)         
&\mapsto      
&\frac{1}{m_{11}}
\left(
\begin{matrix}
1 & -m_{12} \\
m_{21} & \text{det}(M)
\end{matrix}
\right) \\
\end{array}
\end{equation}
This operation $\Pi$ is an involution transforming $\cS$ into $\cT$ as well as $\cT$ into $\cS$~:
\begin{equation}
\cT=\Pi(\cS)\quad,\quad
\cS=\Pi(\cT)\quad,\quad
\Pi\circ\Pi = \mathcal{I}~.
\end{equation}

For a two-step scattering process such as the one discussed here, the full process is described by a  mere
product for $\cT$ matrices that is also by a $\star$ operation for elementary $S$ matrices defined as follows \cite{Genet2003}~:
\begin{equation}
 (\cS_a,\cS_b) \mapsto \cS_a\star \cS_b \equiv \Pi[\Pi(\cS_a)\times \Pi(\cS_b)]~.
\end{equation}
Here, the first step corresponds to reflection on the $\bV_4$ potential, described by a known matrix $\cS_4$, while the second step is the reflection above the tail $\bz<0$ of the real potential $\bV$, that will be discussed below as a matrix $\cS_\rho$ matrix. 
\begin{figure}[ht]
\centering
  \begin{tikzpicture}[scale=1.6]
	  \clip (-1.5,2) rectangle (1.5,-0.75);
	   \draw[myarrow] (-1.5,1) -- node[above]{$a_+^\iin$}(-0.8,1);
	  \draw[myarrow]  (-0.8,0)-- node[above]{$a_-^\out$}(-1.5,0);
	   \draw[fill=gray!40] (-0.6,0.5) ellipse (0.2 and 1);
	     \draw (-0.6,0.5) node{$\cS_\rho$};
	     \draw[myarrow] (-0.35,1) -- (0.35,1);
	  \draw[myarrow]  (0.35,0)-- (-0.35,0);
	  \draw[fill=gray!40] (0.6,0.5) ellipse (0.2 and 1);
	   \draw (0.6,0.5) node{$\cS_4$};
	  \draw[myarrow] (0.8,1) -- node[above]{$a_+^\out$} (1.5,1);
	  \draw[myarrow] (1.5,0) -- node[above]{$a_-^\iin$}(0.8,0);
  \end{tikzpicture}
\caption{Schematic representation of a two-step scattering process.}
\label{scattering}
\end{figure}
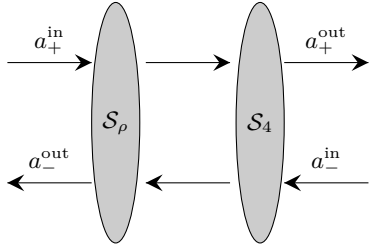 

The whole process, schematized in figure \ref{scattering}, is then described by the $\cS$ matrix~:
\begin{equation}
\cS=\cS_\rho \star \cS_4 \equiv 
\left(
\begin{matrix}
\overline{t}& r \\
\overline{r} & t
\end{matrix}
\right).
\label{global}
\end{equation}
This decomposition might appear as artificial, as the two processes take place in the same space region. 
This argument can however be bypassed as $\cS$ and $\cS_4$ can both be computed numerically and  the $\cS_\rho$ matrix then defined without ambiguity as 
$\cS_\rho=S\star \cS_4^{-1}$
where $\cS_4^{-1}$ is the inverse matrix of $\cS_4$ for the $\star$ law~:
\begin{equation}
\cS_4^{-1}=\Pi([\Pi(\cS_4)]^{-1})
\end{equation}
with $[\Pi(\cS_4)]^{-1}$ the inverse of $\Pi(\cS_4)$ for the usual product law of matrices. 
This gives a proper formal definition of $\cS_\rho$ which matches the physical intuition of a two-step process.
We will see below that this is the key to the improved effective range expansion obtained in this paper.

We first focus attention on the symmetries of the scattering problem,
namely \textit{unitarity} and \textit{reciprocity} which are general symmetries 
and then \textit{space parity} which holds for the specific potential $\bV_4$. 
Unitarity of the $\cS-$matrices associated with current conservation is valid
for all processes considered here~:
\begin{align}
&\cS^\dag \cS=\mathcal{I} ~, \label{unitarity} \\
&|r|^2+|t|^2=|\overline{r}|^2+|\overline{t}|^2=1\quad,\quad
\overline{t}^{*}r+\overline{r}^{*}t=0 ~.\nonumber
\end{align}
Reciprocity is the result of time-reversal symmetry~:
\begin{align}
&\cS_4^{*}=M\cS_4^{-1}M \quad,\quad
M=\left(
\begin{matrix}
0& 1 \\
1 &0
\end{matrix}
\right)~,
\label{reciprocity}\\
&|\text{det}\,\cS|^2=|t \overline{t}-r\overline{r}|^2=1
~.\nonumber
\end{align}
Space parity is a property of the transformed $\bV_4$ potential. It implies that the reflection and transmission amplitudes are the same if we consider a scattered wave coming from the left or from the right~:
\begin{equation}
\label{parity}
\overline{r_4}=r_4\quad,\quad
\overline{t_4}=t_4 ~.
\end{equation}

As a consequence of these symmetries, the $S_4$ matrix has a simple form~:
\begin{equation}
S_4=\left(
\begin{matrix}
t_4& r_4 \\
r_4 &t_4
\end{matrix}
\right)~.
\end{equation}
Its eigenvalues $s_\pm=t_4\pm r_4$ have unit modulus (unitarity) and they can be written as pure dephasings  $s_\pm=e^{i\delta_\pm}$ ($\delta_\pm\in \mathbb{R}$). The known solution of the scattering problem for the $V_4$ potential gives \cite{Dufour2015jpb}~:
\begin{equation}
\begin{split}
r_4&=-i\frac{\sinh(\sigma)}{\sinh(\sigma+i\pi\tau)}\\
t_4&=\frac{\sin(\pi \tau)}{\sinh(\sigma + i\pi \tau)}
\end{split}
\end{equation}
where $\tau$ is a Mathieu characteristic exponent and $\sigma$ the ratio $\ln \frac{\tilde{\psi}^{-}(0)}{\tilde{\psi}^{+}(0)}$, corresponding to the two solutions ($\epsilon=\pm1$) of the Mathieu equation \cite{Dufour2015jpb}~:
\begin{align}
&\tpsi^\epsilon(\tz)=\sum_{n=-\infty}^{\infty}(-1)^nA_n^\tau J_{\epsilon(n+\tau)}(\bk e^{\tz})J_{\epsilon n}(\bk e^{-\tz})
\nonumber \\
&((\tau+2n)^2-1/4)A_n^\tau+\bk^2(A_{n+1}^\tau+A_{n-1}^\tau)=0.
\end{align}

We may now write the reflection amplitude for the full scattering process, using these properties and the reflection amplitude $\rho$ (still to be calculated)~:
\begin{equation}
r=\frac{r_4-\rho \left(r_4^2- t_4^2\right)}{1-\rho r_4}=r_4\frac{1-\rho/r^{*}_4}{1-\rho r_4}~,
\label{rglobal}
\end{equation}
where we have used \eqref{unitarity}, \eqref{reciprocity} and \eqref{parity} to rewrite the determinant of the $S_4-$matrix~:
\begin{equation}
r_4^2-t_4^2=\frac{r_4}{r^{*}_4}~.
\label{rel2}
\end{equation}
We can also reverse the problem and express $\rho$ from expressions of $r$ and $r_4$~:
\begin{equation}
\rho=\frac{r^{*}_4}{r_4}\frac{r_4-r}{1-r^{*}_4r}~.
\label{rho}
\end{equation}

\section{Derivation of the effective range expansion}
We now derive an expansion of $\tA$ which is invariant under the Liouville transformation~:
\begin{equation}
\tA(k)=-i\frac{1+r(k)}{1-r(k)}
=-i\frac{1+\br(\bk)}{1-\br(\bk)}=\tA(\bk)~.
\label{acal}
\end{equation}

The expansion of $\tA_4$ is known \cite{OMalley1961}~:
\begin{align}
\tA_4(k)&=-ik\ell_4\left[\alpha_0+\alpha_1 k\ell_4+\alpha_2 (k\ell_4)^2 \right.
\nonumber \\
&\qquad\qquad \left.+\alpha_2^\prime(k\ell_4)^2\ln k\ell_4]+\mathcal{O}(k^4 )\right.~,
\nonumber \\
\alpha_0 &= 1 \quad,\quad  \alpha_1 = \frac{\pi}{3}i  \quad,\quad \alpha_2^\prime=\frac{4}{3}~,
\nonumber \\
\alpha_2&=\frac{8}{3}(\gamma+\ln 2)-\frac{28}{9}-\frac{2\pi}{3}i~,
\label{alphai}
\end{align}
and that of $\tA_4(\bk)$ deduced from \eqref{rho} ($\bk=\sqrt{k\ell_4}$)~:
\begin{align}
\tA_4(\bk)=&-i\bk^2\left[\alpha_0+\alpha_1 \bk^2+\alpha_2 \bk^4+2\alpha_2^{'}\bk^4\ln \bk\right]
\nonumber \\
&+\mathcal{O}(\bk^8)~.
\label{tA4bk}
\end{align}

Using \eqref{rglobal} and \eqref{acal}, we deduce~:
\begin{equation}
\tA=\tA_4+\frac{\rho(1-i\tA_4)(\tA_4^{*}-\tA_4)}{1+i\tA_4^{*}+\rho(1-i\tA_4)}~.
\label{equA}
\end{equation}
It is interesting to note that if $\tA_4(k) \in \mathbb{R}$, then $\tA=\tA_4$ independently of the value of $\rho$.
We see from \eqref{tA4bk} that we need an expansion of $\tA(\bk)$ up to order 6 in $\bk$. if we want an expansion of $\tA(k)$ up to order 3 in $k$. Since $\tA_4^{*}-\tA_4\sim \bk^2$, this aim only requires an expansion of $\rho(\bk)$ up to the order 4 in $\bk$. 

As the difference $\bV-\bV_4$ is regular and decreases fast enough to apply Lippmann-Schwinger equations in scattering theory, it is natural to postulate that $\rho$ has a regular  Taylor expansion in $\bk$~:
\begin{equation}
\rho(\bk)=\rho_0+\rho_1 \bk + \rho_2 \bk^2+\rho_3 \bk^3 + \rho_4 \bk^4 + \ldots
\label{rho}
\end{equation}
This postulate can then be checked out and the coefficients $\rho_i$ deduced through a fit using 
the exact expression known for $\br_4(\bk)$ and the numerical results of $\br(\bk)$ calculated for different surfaces. 

We have performed these fits in the range $k\ell\in [2\times10^{-3},10^{-1}]$. The numerical noise is indeed too large for $k\ell<2\times10^{-3}$, while the truncated Taylor expansion ceases to be valid for $k\ell>10^{-1}$. We have taken uniformly 1000 points in the interval to build the discreet set of points to be fitted. The results of this fit are presented in table \ref{coeffs}.

\begin{table}[ht]
\begin{tabular}{c|c|c|}
\cline{2-3}
 & He & SiO$_2$  \\
\hline
\multicolumn{1}{|c|}{$\rho_0$} & $0.158+0.336i$ & $0.064+0.138i$ \\
\hline
\multicolumn{1}{|c|}{$\rho_1$} & $-0.009+0.011i$ & $-0.001-0.004i$ \\
\hline
\multicolumn{1}{|c|}{$\rho_2$} & $0.098-0.117i$ & $0.026+0.034i$ \\
\hline
\multicolumn{1}{|c|}{$\rho_3$} & $-0.513+0.611i$ & $-0.359+0.469i$ \\
\hline
\multicolumn{1}{|c|}{$\rho_4$} & $-0.083-0.487i$ &  $0.204-0.740i$ \\
\hline
\end{tabular}
\caption{Coefficients in the expansion of $\rho$ deduced by fitting the numerical results known for He and SiO$_2$ surfaces. }
\label{coeffs}
\end{table}

This polynomial expansion in $\bk$ introduces square root terms in the original expansion in $k$. Such terms would have been absent in a naive approach keeping only integer powers of $k$, that is equivalently even powers of $\bk$ in the expansion \eqref{rho}. We may compare the quality of the fits corresponds to the \textit{full} Taylor expansion-- truncated after the fourth order --  and to \textit{even terms} expansion restricted to even powers of $\bk$ by looking at the standard deviations of the residuals. These standard deviations $\hat{\sigma}$ calculated for real and imaginary parts of $\rho$, and different surfaces (He or SiO$_2$), are presented in table \ref{sd}. 

\begin{table}[ht]
\centering
\begin{tabular}{c|c|c|c|c|}
\cline{2-5}
\multirow{2}{0.5cm}{$10^{6}~\hat{\sigma}$} & \multicolumn{2}{c|}{He} & \multicolumn{2}{c|}{SiO$_2$}\\
 \cline{2-5}
& Re($\rho$) & Im ($\rho$) & Re($\rho$) & Im ($\rho$) \\
\hline
\multicolumn{1}{|c|}{Even terms expansion} & $31$ & $36$ & $41$ & $80$ \\
\hline
\multicolumn{1}{|c|}{Full expansion} & $2$ &  $2.4$ & $1.0$ & $5.5$ \\
\hline
\end{tabular}
\caption{Standard deviations $\hat{\sigma}$ of the residuals in the fits, calculated for real and imaginary parts of $\rho$, for different surfaces, as a function of the fit method -- based on the even terms and the full expansions.}
\label{sd}
\end{table}

\begin{figure}[ht]
   \center
   \includegraphics[width=0.7\linewidth]{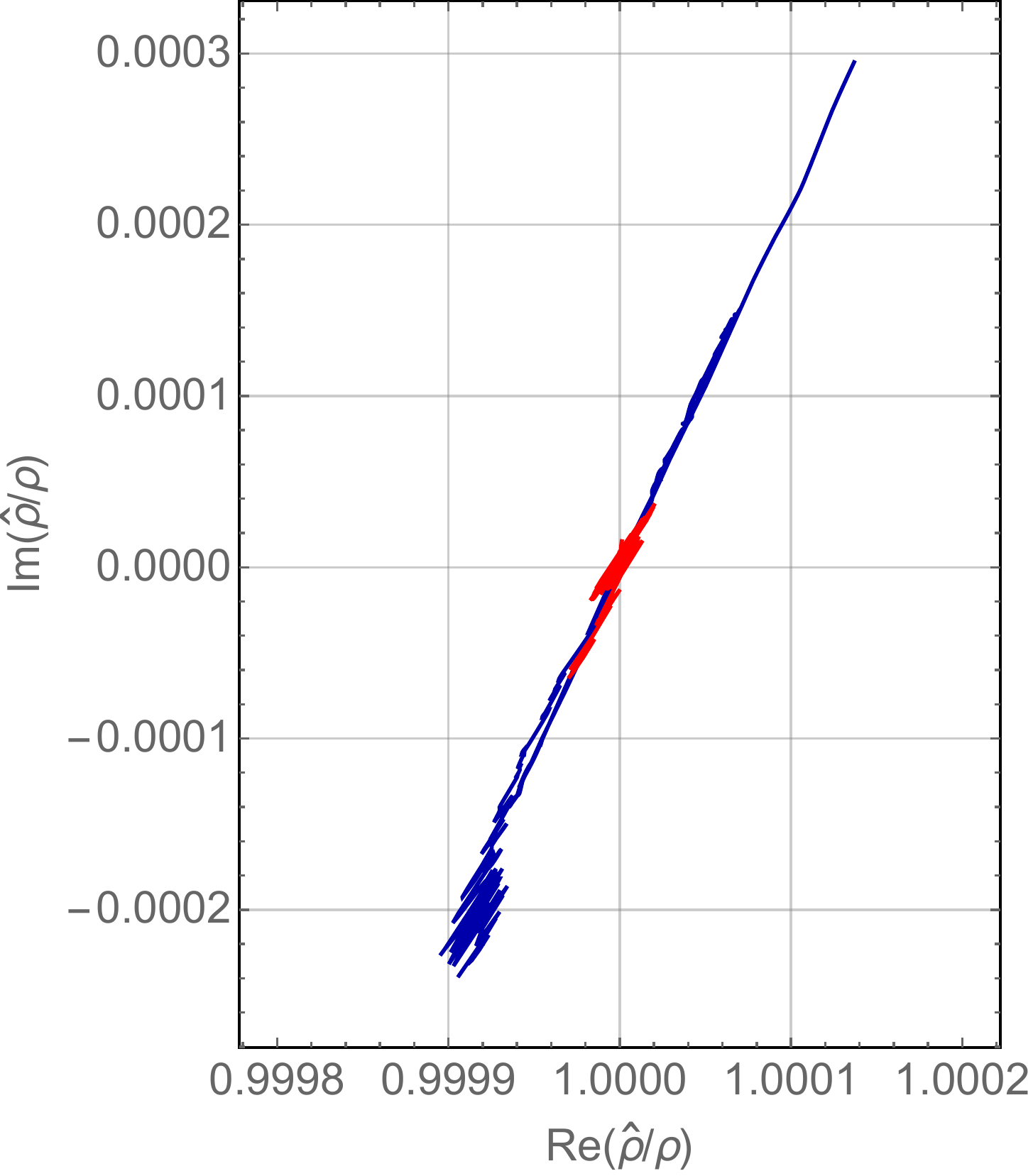}
   \includegraphics[width=0.7\linewidth]{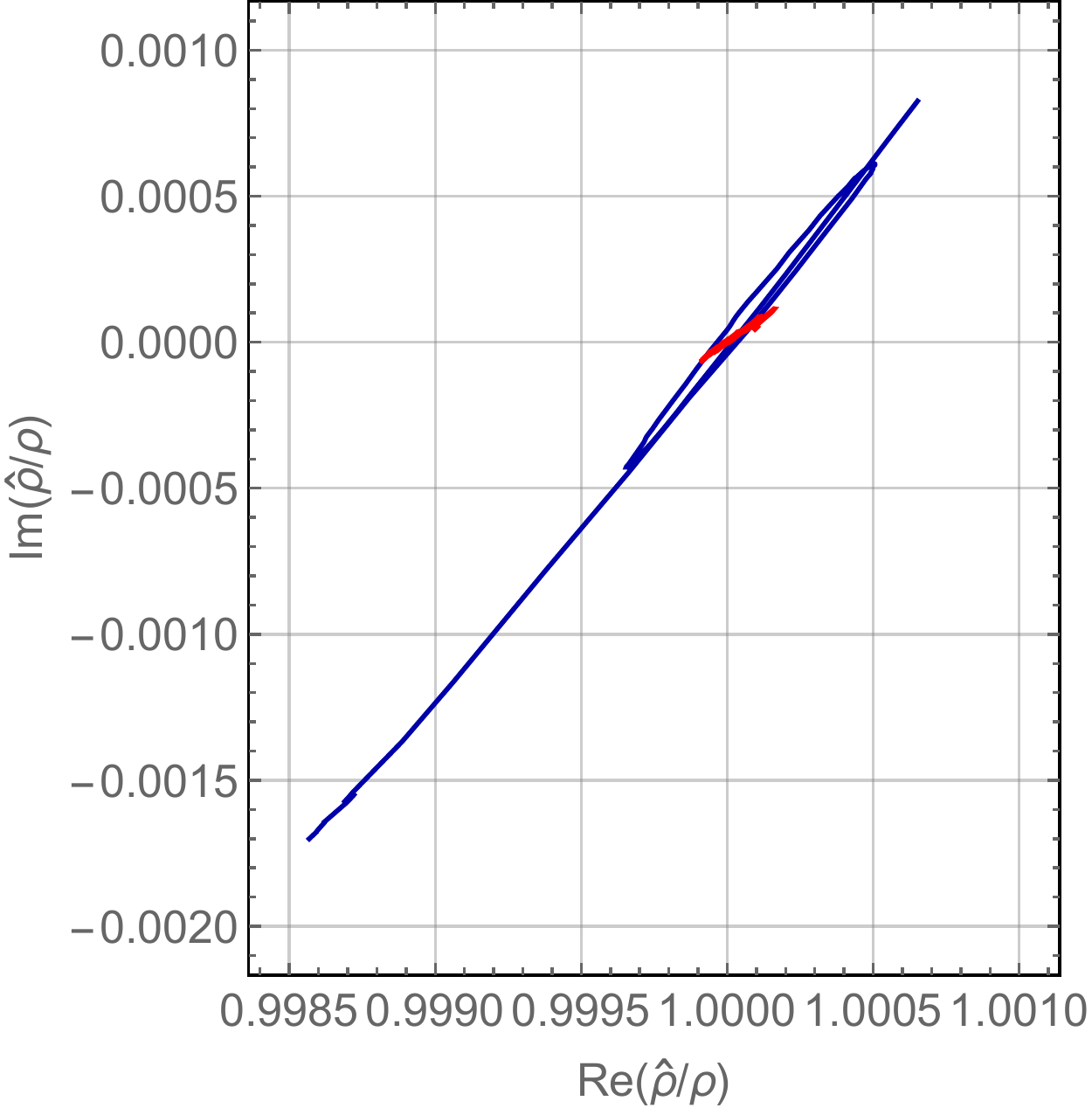}
   \caption{Ratio $\hat{\rho}/\rho$ of the $\hat{\rho}$ obtained by the fit over the $\rho$ known numerically. The upper figure is a plot for He and the right for SiO$_2$. The fit obtained with only even terms is plotted in blue, while the fit obtained with the full Taylor expansion is plotted in red. }
   \label{ratio}
\end{figure}

The ratio $\hat{\rho}/\rho$ of the estimated $\hat{\rho}$ for both even terms and full expansions to the numerically known $\rho$ is another illustration of the quality of the two fits  plotted in figure \ref{ratio}. Both table \ref{sd} and figure \ref{ratio} show a large improvement of the quality of the fit with the full Taylor expansion. The residual linear behavior in figure \ref{ratio} reveals missing terms in the expansion of $\rho$ -- the odd terms in the case of the even terms expansion and terms from the order 5 in the case of the full expansion. Liouville transformations thus reveal the existence of square root terms which could not have been discovered otherwise and neatly improve the precision of the estimators. 

We can finally derive the expansion of $\tA$ knowing the ones for $\tA_4$ and $\rho$~:
\begin{align}
\tA(k)&=-ik\ell \left[\beta_0+\beta_{12}(k\ell_4)^{1/2}+\beta_1 k\ell_4 \right. \nonumber\\
&\qquad+\beta_{32}(k\ell_4)^{3/2}+\beta_2 (k\ell_4)^2 \nonumber\\
&\qquad\left.+\beta_2^{'}(k\ell_4)^2\ln k\ell_4\right]+\mathcal{O}(k^4) ~,
\label{expansion}\\
\ell&=\frac{1-\rho_0}{1+\rho_0}\,\ell_4 ~. 
\end{align}
The last equation is a renormalization of the scattering length $\ell$  differing from the long-range length scale $\ell_4$ due to the non vanishing value of $\rho_0$. 
The other constants in the expansion are deduced from $\alpha_i$ in \eqref{alphai} and $\rho_i$~:
\begin{align}
  \beta_0 =&\alpha_0~, \nonumber\\
  \beta_{12}=&-\frac{2\alpha_0 \rho_1}{\sqrt{(1+\rho_0)(1-\rho_0)^3}}~, \nonumber\\
  \beta_1 =& \frac{2\alpha_0(\rho_1^2-(1+\rho_0)\rho_2)-\alpha_1(1+\rho_0)^3}{(1+\rho_0)(1-\rho_0)^2}~, 
\nonumber\\
  \beta_{32}=&\frac{2\alpha_0\left(\rho_1^3-2(1+\rho_0)\rho_1\rho_2+(1+\rho_0)^2\rho_3\right)}{\sqrt{(1+\rho_0)^3(1-\rho_0)^5}}~, \nonumber\\
  \beta_2=&\frac{2\alpha_0\rho_1^2\left(\rho_1^2-3(1+\rho_0)\rho_2\right)}{(1+\rho_0)^2(1-\rho_0)^3} \nonumber\\
  +&\frac{2\alpha_0(2\rho_1\rho_3+\rho_2^2)}{(1-\rho_0)^3} \nonumber\\
+&\frac{1+\rho_0}{(1-\rho_0)^3} (\alpha_2+4\alpha_0\alpha_1\rho_0-2\alpha_0\rho_4)
\nonumber\\
  +&\frac{\rho_0(1+\rho_0)}{(1-\rho_0)^3} ((2+\rho_0)\text{Im}(\alpha_2)i-\rho_0\text{Re}(\alpha_2))
\nonumber\\
   -& \frac{(1+\rho_0)^2}{(1-\rho_0)^2} \alpha_2^\prime \ln \left(\frac{1-\rho_0}{1+\rho_0} \right) ~,
\nonumber\\
  \beta_2^\prime=&\alpha_2^\prime \left(\frac{1+\rho_0}{1-\rho_0}\right)^2~.
\label{betai}
\end{align}

\section{Discussion of the new expansion}

We conclude our discussion of the improved effective range expansion by showing its benefit with respect to the modified effective range theory \cite{OMalley1961}. 

In the latter, two parameters had to be modified in the expansion of $\tA$: the scattering length $\ell$ and the effective range $\mathcal{R}_0$, or equivalently $\tilde{\alpha}_2$~:
\begin{align}
&\tA(k)=-ik\ell \left(\tilde{\alpha}_0+\tilde{\alpha}_1 k\ell_4+\tilde{\alpha}_2 (k\ell_4)^2\right.\nonumber\\
&\qquad\qquad\left.+\tilde{\alpha}_2^\prime(k\ell_4)^2\ln k\ell_4\right) +\mathcal{O}(k^4) ~, \nonumber \\
&  \tilde{\alpha}_0 =\alpha_0 \quad,\quad \tilde{\alpha}_1 = \alpha_1\cdot\ell_4/\ell \quad,\quad
\tilde{\alpha}_2^\prime=\alpha_2^\prime ~,  \nonumber \\
&  \tilde{\alpha}_2=\alpha_2+\pi(\ell-\ell_4)^2/\ell\ell_4-i\mathcal{R}_0\ell/2\ell_4^2 ~,
\label{Aktilde}
\end{align}
with the numerical values of the scattering length $\ell$ and $\tilde{\alpha}_2$ for helium and silica in table \ref{aandalpha2}.
\begin{table}[htp]
\centering
\begin{tabular}{c|c|c|}
\cline{2-3}
 & He & SiO$_2$  \\
\hline
\multicolumn{1}{|c|}{$\ell$} & $44.78-34.90i$ & $272.7-77.04i$ \\
\hline
\multicolumn{1}{|c|}{$\tilde{\alpha}_2$} & $2.54-2.51i$ & $1.73-3.67i$ \\
\hline
\end{tabular}
\caption{Modified values for the scattering length $\ell$ and the coefficient $\tilde{\alpha}_2$ computed for helium and silica surfaces.}
\label{aandalpha2}
\end{table}

In the improved effective range expansion proposed in this paper, five parameters have to be determined, $\rho_i$ for $i\in \llbracket 0,4 \rrbracket$, some of them leading to new terms in the expansion of $\tA$. We calculate the error due to the two expansions by comparing the values of the expansions to the precise numerical one. 
\begin{figure}[htp]
   \center
   \includegraphics[width=0.70\linewidth]{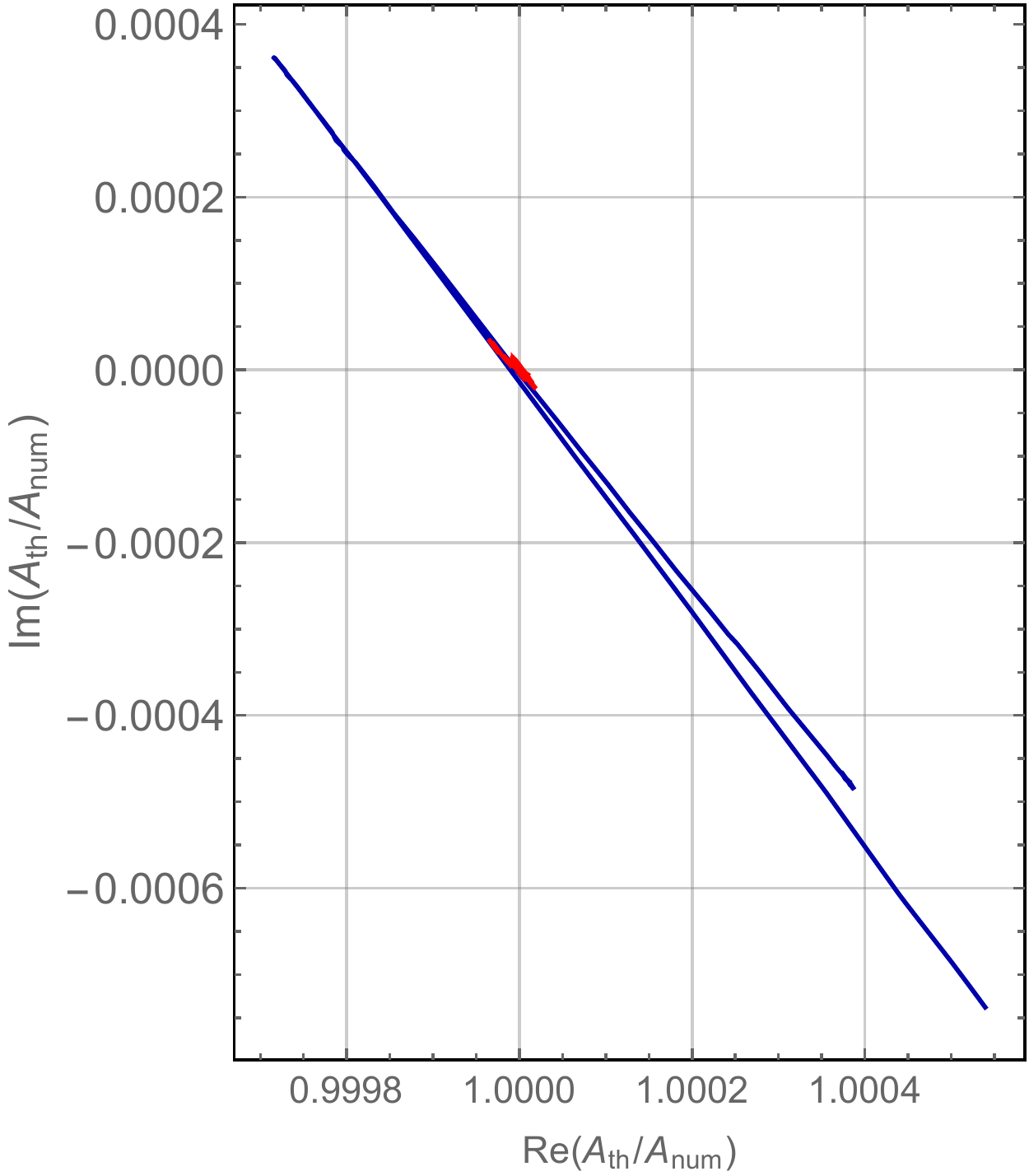}\quad \quad \quad
   \includegraphics[width=0.68\linewidth]{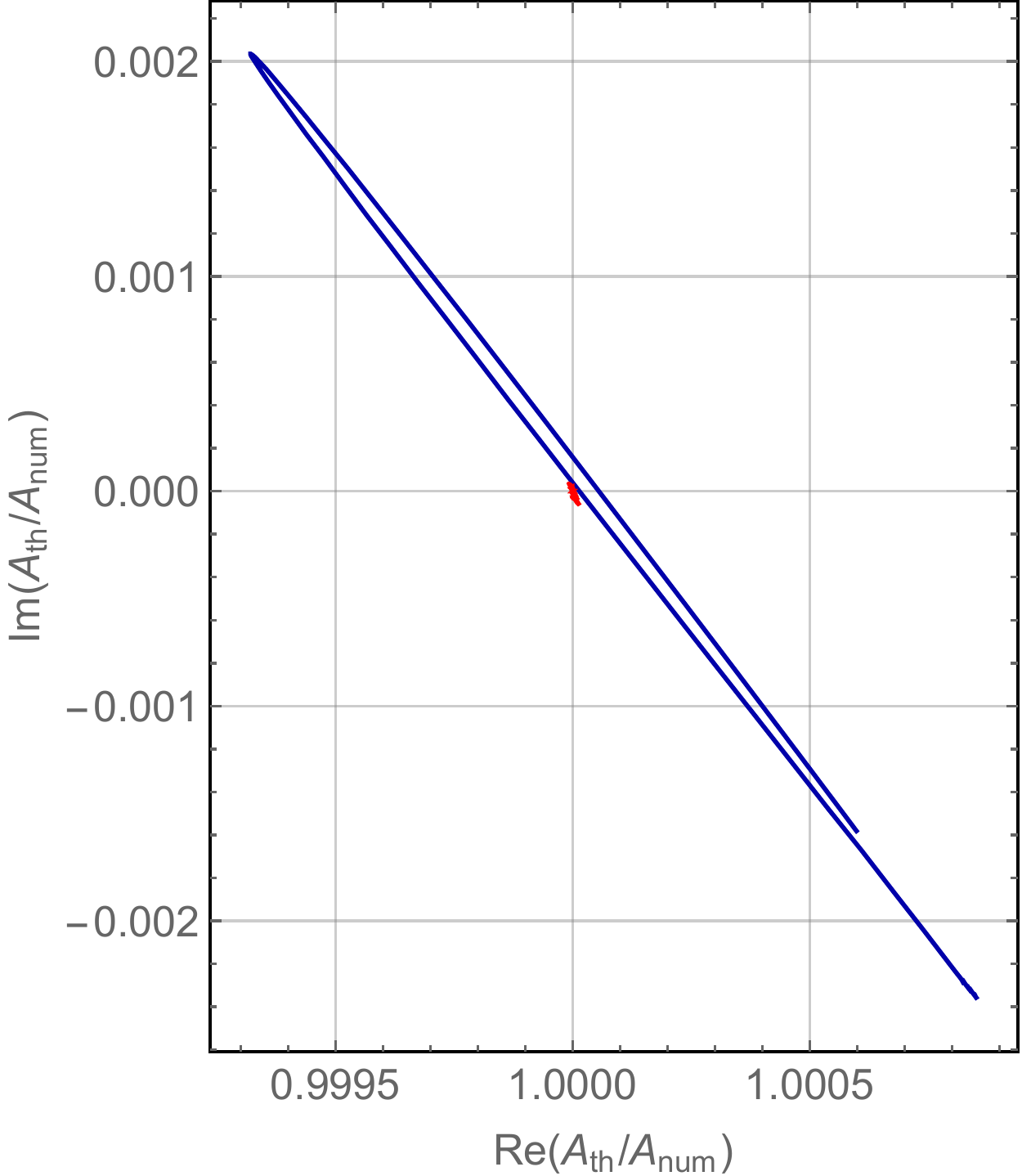}
   \caption{Ratio $\tA_{th}/\tA_{num}$ of $\tA_{th}$ obtained theoretically over $\tA_{num}$ known numerically. The modified effective range theory appears in blue while the improved effective range theory appears in red. The upper figure is a plot for He and the lower one for SiO$_2$.  }
   \label{ratioA}
\end{figure}

We plotted in figure \ref{ratioA} the ratio $\tA_{th}/\tA_{num}$ of $\tA_{th}$ in the complex plan obtained theoretically for the preceding effective range theory and the improved one, over $\tA_{num}$ known numerically. It is clear that the improved effective range theory reproduces the energy dependance of $\tA$ with a much better accuracy than the original theory. 
To be more quantitative, we define a norm as the maximum deviation of the predicted and numerically known values~:
\begin{equation}
\label{relativeerror}
\left | \left | f \right | \right | \equiv \underset{k\ell_4 \in [2\cdot 10^{-3},10^{-1}]}{\max} \;  \left | f(k\ell_4) \right | ,
\end{equation}
and compare $\left | \left |\frac{\tA_{th}}{\tA_{num}} -1\right | \right |$  for the two theories in table \ref{relativeA}. These results confirm what was already visible in the figure \ref{ratioA}. The new improved expansion is more than 10 times more accurate than the expansion derived from the original effective range theory. 
\begin{table}[htp]
\centering
\begin{tabular}{c|c|c|}
\cline{2-3}
 & He & SiO$_2$  \\
\hline
\multicolumn{1}{|c|}{Modified ERT} & $9.1\times 10^{-4}$ & $2.5\times10^{-3}$ \\
\hline
\multicolumn{1}{|c|}{Improved ERT} & $4.8\times10^{-5}$ & $4.3\times10^{-5}$ \\
\hline
\end{tabular}
\caption{Maximum relative error for the old effective range theory and the improved effective range expansion for helium and silica surfaces.}
\label{relativeA}
\end{table}

The improvement is partly due to the fact that the improved expansion  allows more degrees of freedom than the old one (5 instead of 2). But the key improvement is the addition of new terms  in the expansion \eqref{expansion} which  represent a much better physical understanding of the two-step  scattering process. These terms come from the better defined scattering problem after Liouville transformation. The coefficients $\rho_i$ obtained with the new fit are sufficient to compute the expansion of $\tA$ with a very high accuracy, needed to determine precisely the Casimir-Polder shifts of quantum gravitation states \cite{Crepin2017} and, then, to take full benefit of the recently proposed quantum interference method \cite{Crepin2019}.  

\emph{Aknowledgments }
Thanks are due to G. Dufour, A. Lambrecht, V. Nesvizhevsky and A. Voronin for insightful discussions during this work
 and also to colleagues in the GBAR and GRANIT collaborations.


\end{document}